\documentclass[pra,twocolumn,aps,superscriptaddress]{revtex4} %
\usepackage{dsfont}
\newcommand{\tr}{\mathrm{tr}\,}
\newcommand{\comment}[1]{}

\begin{document}
\bibliographystyle{apsrev}

\title{Secure Controlled Teleportation}
\author{Dan Kenigsberg}
\email{danken@cs.technion.ac.il}
\author{Tal Mor}
\affiliation{Computer Science Department, Technion, Haifa 32000, Israel.}

\begin{abstract}
Several protocols for controlled teleportation were suggested by
Yang, Chu, and Han [PRA \textbf{70}, 022329 (2004)]. In these protocols, 
Alice teleports qubits (in an unknown state) to Bob iff a controller allows it.
We view this problem in the perspective of secure multi-party quantum
computation.
We show that the suggested entanglement-efficient protocols for
$m$-qubit 
controlled teleportation 
are open to cheating; Alice and Bob may teleport
$(m-1)$-qubits of quantum information, \textbf{out of the controllers' control.}
We conjecture that the straightforward protocol for controlled teleportation,
which requires each controller to hold $m$ entangled qubits, is 
optimal. We prove this conjecture for a limited, but interesting, subset of
protocols.
\end{abstract}

\maketitle 

\def\bra#1{\left<#1\right|}
\def\ket#1{\left|#1\right>}
\def\avg#1{\langle#1\rangle}

\section{Introduction}
Secure multi-party quantum computation (MPQC) protocol~\cite{CGS02} 
allows $n$ players to compute an agreed quantum circuit
where each player has access only to his own arbitrary quantum input.
A MPQC protocol has two phases: In the \emph{sharing} phase, players dubbed
dealers provide the other players with their initial state.  In the
\emph{reconstruction} phase, the honest players help a designated player 
reconstruct the final state of the protocol. During the latter phase, only local
operations and classical computation is available.

In this paper we view \emph{controlled teleportation} as a special case of
MPQC. A dealer named Carol hands Alice and Bob an (entangled) initial state $\ket{\psi_{ABC}}$.
A second dealer, named David, provides Alice with an unknown $m$-qubit state
$\rho$. 
The task of Alice and Bob is to reconstruct (teleport) $\rho$
into Bob's hands, when Carol allows it. 
$\ket{\psi_{ABC}}$ is such that Carol controls whether the teleportation
can take place.
Carol and David are honest dealers.
We call a controlled teleportation protocol \emph{secure} if it is impossible
for malicious Alice and Bob to teleport any part of $\rho$ before the
reconstruction phase. Namely, if Alice and Bob can build a state $\rho'$ at
Bob's hands which has non-trivial fidelity with $\rho$ before the
reconstruction phase, the protocol is insecure.

The straightforward solution to this problem, suggested in~\cite{YangChuHan04}, 
is to use a procedure described 
in~\cite{KB98,HBB99}.
First, Carol prepares the following $3m$-qubit state and gives Alice and Bob
their respective qubits
\begin{eqnarray}
&&\otimes_{i=1}^m\ket{GHZ}_{ABC(i)}=\nonumber\\
&&\otimes_{i=1}^m\left(\ket{\phi^+}_{AB(i)}\ket+_{C(i)}+
\ket{\phi^-}_{AB(i)}\ket-_{C(i)}\right).
\end{eqnarray}
(Here, and throughout the paper, we drop normalization factors for
readability), $\{\ket{\psi^\pm},\ket{\phi^\pm}\}$ are the four
Bell-BMR\cite{BMR92} states,
$\ket{\pm}=\frac{\ket0\pm\ket1}{\sqrt2}$, and
$\ket{GHZ}_{ABC(i)}=\frac{\ket{000}+\ket{111}}{\sqrt2}$ is the $i$th
$GHZ$ state shared among Alice, Bob and Carol.
Later, if Carol wishes to allow the teleportation, she measures her $m$ qubits
in
the Hadamard basis, and publishes her results $c_i\in\{+,-\}$.
Now, the state shared by Alice and Bob is $\otimes_{i=1}^m\ket{\phi^{c_i}}$
which can be freely used by them for teleportation.

On the other hand, if Carol abstains from participation, the state shared by
Alice and Bob can be calculated by tracing over Carol's qubits
\begin{eqnarray}
&&\tr_C\otimes_{i=1}^m\ket{GHZ}\bra{GHZ}_{ABC(i)}=\nonumber\\
&&\otimes_{i=1}^m\left(\ket{00}\bra{00}_{AB(i)}+\ket{11}\bra{11}_{AB(i)}\right).
\end{eqnarray}
We note here, that without the participation of Carol,
the state shared by Alice and Bob becomes a classical correlation,
which cannot facilitate quantum teleportation.

Ref.~\cite{YangChuHan04} provides a second protocol, in which Carol holds only
one entangled qubit, aiming at the
same task. This entanglement-efficient protocol can be stated as follows: 
Carol creates the following $2(m+1)$-qubit state and
gives Alice and Bob their respective qubits
\begin{eqnarray}
\otimes_{i=1}^m\ket{\phi^+}_{AB(i)}\otimes\ket{\phi^+}_{AC}+
\otimes_{i=1}^m\ket{\phi^-}_{AB(i)}\otimes\ket{\psi^+}_{AC}.
\end{eqnarray}
Later, if Carol wishes to allow the teleportation, she measures her single qubit
in the computational basis and publishes her result. 
Alice measures her own rightmost qubit in the computational basis. If it is
equal to Carol's outcome, then Alice and Bob share
$\otimes_{i=1}^m\ket{\phi^+}_{AB(i)}$. Otherwise, they share
$\otimes_{i=1}^m\ket{\phi^-}_{AB(i)}$. Either way, they can safely teleport $m$
qubits.
On the other hand, if Carol abstains from participation, and 
if Alice and Bob continue the protocol exactly as planned, they can no longer teleport
Alice's $m$-qubit message reliably, since they will create a mixed
state~\cite[Eq. (11)]{YangChuHan04} instead.

\section{Alice and Bob can cheat Carol}
In fact, in the second protocol, 
even if Carol does not participate,
malicious Alice and Bob can let Alice teleport any $(m-1)$-qubit state to Bob.
The abstention of Carol mixes the shared state to create 
\begin{eqnarray}
\otimes_{i=1}^m\ket{\phi^+}\bra{\phi^+}_{AB(i)}\otimes\mathds1_{A}+
\otimes_{i=1}^m\ket{\phi^-}\bra{\phi^-}_{AB(i)}\otimes\mathds1_{A},
\label{prot2noCarol}
\end{eqnarray}
where $\mathds1$ is the totally mixed state in one qubit.
Yet Alice and Bob
can easily distill it~\cite{BBPSSW96}.
Each of them has to relinquish his or her $m$th qubit
and measure it in the computational basis.
If their results coincide, they share $\otimes_{i=1}^{m-1}\ket{\phi^+}_{AB(i)}$;
otherwise they share $\otimes_{i=1}^{m-1}\ket{\phi^-}_{AB(i)}$. Either way, they can
safely teleport  any $(m-1)$-qubit state and thus reconstruct any $m$-qubit
state with high fidelity.

It is important to note that~\cite{YangChuHan04} 
never claimed that Bob can learn \emph{nothing} about Alice's state,
and they stated openly that they did not ``attempt a comprehensive study of the
security against all  possible forms of eavesdropping and/or cheating''.
However, it is equally important to note that their efficient protocol for
multiqubit quantum information teleportation via the control of an agent, is
insecure.
The same malady affects their protocol for multiple controllers (see
Section~\ref{multicontrol}).

\section{Carol needs $m$ entangled qubits}
When Carol held a single qubit entangled to Alice
and Bob, she could control only one of their qubits, and not the rest.
We believe that this is not an accident. We \emph{conjecture} that
Carol must have at least $m$ entangled qubits with Alice and Bob if she wants
to completely control their ability to teleport an $m$-qubit state.

We prove a special case of this conjecture.
We define a limited form of secure controlled
teleportation. In the limited form, 
we assume three additional limitations. 
(a) The initial state shared by Alice, Bob and Carol is pure.\footnote{In
general, this state could have been mixed.}
(b) If Carol abstains, the remaining state
$\rho_{AB}=\tr_C\ket{\psi_{ABC}}\bra{\psi_{ABC}}$ is separable.\footnote{In
general, it is probably
enough to assume that the state without Carol is not distillable, namely
either separable or bound-entangled.}
(c) In the reconstruction phase, Carol performs her measurement on the shared
state without obtaining any prior information from Alice and Bob; Alice and Bob
do not help Carol to assist them.\footnote{In general, Carol's measurements can
depend on the outcome of Alice and Bob's measurements; The reconstruction phase
can be more complex.}

These limitations are not true in general, since the initially-shared state may
be mixed, since a bound-entangled state is probably equally
unhelpful for Alice and Bob if they want to perform teleportation,
and since the reconstruction phase can be more complex.
Note that these limitations leave enough room for interesting protocols.
Specifically, the protocols of~\cite{YangChuHan04} satisfy limitations (a)
and (c), and the ones not satisfying (b) can be cheated because of that.

Limitation (c) 
means that the highest value of entanglement that Alice
and Bob can create between them with the help of Carol is $EoA^1(\rho_{AB})$,
the entanglement of assistance, which in turn is limited by $EoA^\infty(\rho_{AB})$.
Recently, Smolin, Verstraete, and
Winter~\cite{SVW05} showed that
\begin{equation}
EoA^\infty(\rho_{AB})\leq \min(S(A),S(B)).
\end{equation}

Let us now assume that a secure limited controlled teleportation protocol exists,
i.e.\ there exists $\ket{\psi_{ABC}}$ so that
$\rho_{AB}=\tr_C{\ket{\psi_{ABC}}\bra{\psi_{ABC}}}$ is separable, while
$EoA^1(\rho_{AB})\geq m$.
Since $\rho_{AB}$ is separable,
$S(\rho_{AB})\geq \max(S(A),S(B))$.
We conclude that
\begin{eqnarray}
S(\rho_{AB})&\stackrel{\hbox{(b)}}{\geq} &\max(S(A),S(B))\nonumber\\
&\geq& \min(S(A),S(B)) 	\stackrel{\hbox{\cite{SVW05}}}\geq
EoA^\infty(\rho_{AB})\\
&\geq&EoA^1(\rho_{AB})\stackrel{\hbox{(c)}}\geq m.\nonumber
\end{eqnarray}
But since $\ket{\psi_{ABC}}$ is pure, $S(\rho_{AB})$ is exactly the initial
entanglement of Carol with Alice and Bob. 
Thus, a limited controlled teleportation protocol requires Carol to hold no less
than $m$ entangled bits---just as in the straightforward protocol.

\section{Multiple controllers}
\label{multicontrol}
In an extended problem presented in~\cite{YangChuHan04}, Carol is replaced by
$n$ controllers: If all the controllers participate, 
Alice can teleport an $m$-qubit
message to Bob. But even if a single controller abstains, the teleportation has
to be impossible.

In the straightforward protocol achieving this,
the controllers prepare the following $(n+2)m$-qubit state and give
each participant his or her respective qubits
\begin{eqnarray}
&&\otimes_{i=1}^m\ket{GHZ}_{ABC^n(i)}=\nonumber\\
&&\otimes_{i=1}^m\left(\ket{\phi^+}_{AB(i)}
      H^{\otimes n}\sum_{\mathrm{even}\,|x|}\ket x_{C^n(i)}+\right. \nonumber\\
&&~~~~~~~~\left.\ket{\phi^-}_{AB(i)}H^{\otimes n}\sum_{\mathrm{odd}\,|x|}\ket x_{C^n(i)}\right)
\end{eqnarray}
where $H^{\otimes n}$ is the Hadamard transform on $n$ qubits, and
the summations are over the
$n$-bit strings $x$ whose Hamming weight $|x|$ is even (odd).
Later, when all the controllers wish to allow the teleportation, each of them
apply the Hadamard transform to her first qubit, 
measures it in the computational basis, 
and publishes her result. 
If the number of ``1''s published by all 
controllers is even (odd), Alice and Bob share $\ket{\phi^+}$ ($\ket{\phi^-}$).
Either way, they can safely teleport one qubit. 
This is repeated on the $m-1$ sets of remaining qubits.

On the other hand, if even a single controller abstains, the complete state
becomes a classical correlation, unworthy for teleportation.

In this protocol each controller initially holds $m$ entangled bits with the
rest of the system. Much like as in the case of a single controller, the
protocol cannot be improved by another protocol of the limited form.

If all $n$ controllers participate,
they can be thought of as one, and $m\leq EoA(\rho_{AB})\leq \min(S(A),S(B))$.

If one of the controllers $(C')$ abstains, we require that
$\rho_{ABC^{n-1}}$ would become separable. Again, this means that each
controller's entanglement with the rest of the system 
$S(C')=S(\rho_{ABC^{n-1}})$ has to be more than $\max(S(A),S(B))$
and certainly more than $\min(S(A),S(B))\geq m$.

Any limited controlled teleportation protocol that tries to
be more entanglement-efficient than that, is insecure. 
For example, in the shared state suggested in~\cite[Eq.~(21)]{YangChuHan04}
\begin{eqnarray}
\otimes_{i=1}^m\ket{\phi^+}_{AB(i)}\otimes\ket{GHZ_+}_{AC^n}+\nonumber\\
\otimes_{i=1}^m\ket{\phi^-}_{AB(i)}\otimes\ket{GHZ_-}_{AC^n}
\end{eqnarray}
(where $\ket{GHZ_\pm}=\ket{0...0}\pm\ket{1...1}$ is an $n+1$ qubit state share
by Alice and the $n$ controllers) there is only one bit of entanglement
between the group of all controllers and the Alice and Bob pair.
Therefore, Alice and Bob can again ignore the controllers and remain
with a state useful for teleportation.

\section{Acknowledgments}
We thank Amir Kalev and Gili Bisker for discussing this paper with us 
and for their valuable comments, and for the support of 
the Israeli MOD Research and Technology Unit.




\end{document}